\documentclass[journal]{IEEEtran}
\usepackage{amsmath,amsfonts}
\usepackage{algorithmic}
\usepackage{algorithm}
\usepackage{array}
\usepackage[caption=false,font=normalsize,labelfont=sf,textfont=sf]{subfig}
\usepackage{textcomp}
\usepackage{stfloats}
\usepackage{url}
\usepackage{verbatim}
\usepackage{graphicx}
\usepackage{xcolor}
\usepackage{cite}
\usepackage{bm}
\usepackage{booktabs}
\usepackage{hyperref}

\newcommand{\uop}{\textmu{}Op}

\begin{document}
\bstctlcite{IEEEexample:BSTcontrol}  %

\title{Hidden in Plain Sight: Privacy-Enhanced Computing Through Register Renaming}

\author{Ahmed Nematallah and Stefanos Kaxiras, \IEEEmembership{Fellow, IEEE}

\thanks{Ahmed Nematallah and Stefanos Kaxiras are with the Computer Science Department at Uppsala University (e-mails: ahmed.nematallah@it.uu.se, stefanos.kaxiras@it.uu.se). }
\thanks{This work was supported by the Knut and Alice Wallenberg Foundation (KAW) through the Wallenberg Scholars Prolongation program (grant no. 2020.0202), the Swedish Research Council VR grant 2022-04959, and the Swedish Foundation for Strategic Research (SSF) grant FUS21-0067}}

\markboth{Preprint}%
{Hidden in Plain Sight: Privacy-Enhanced Computing Through Register Renaming}

\maketitle
\begin{abstract}
Privacy-Enhanced Computing (PEC) is a mechanism to hide program secrets from the outside world without paying the cost of fully-homomorphic encryption. 
Previous work has developed hardware mechanisms to achieve this goal but they require a dedicated processor enclave where all computations are done, requiring duplicate ALU resources and limiting the performance of PEC-utilizing applications due to not being able to access the full compute resources of the processor (e.g., multiple execution units and vector units), or not being able to execute PEC instructions in parallel and out-of-order.

In this work, we introduce HPS-PEC (Hidden in Plain Sight PEC), a novel method of utilizing existing infrastructure in the CPU to achieve PEC with minor modifications to the architecture.

We show that, across 25 VIP-Bench benchmarks and with a 7-cycle QARMA-64 cipher, HPS-PEC runs encrypted code at 3.41$\times$ the execution time of unprotected native code (geometric mean), of which 1.03$\times$ is the cost of data-oblivious code itself, and is 1.26$\times$ faster than a Sequestered Encryption (SE) unit while utilizing under a third of its dedicated area. Our method keeps plaintext out of reach of every instruction, as an enclave does given a correctly functioning renamer and register file, which every program already depends on.
\end{abstract}


\section{Introduction}

Modern computing systems routinely process sensitive data, but standard methods require data to be decrypted before it can be computed on, exposing it to whoever runs the computation. Privacy-Enhanced Computing (PEC) addresses this problem by allowing computations to be performed directly on encrypted data, so the underlying information is never exposed. Several approaches have been developed to achieve this, including homomorphic encryption, secure multiparty computation, trusted execution environments, and sequestered encryption (SE). Sequestering in particular has garnered attention lately~\cite{biernacki2021vip}, where sensitive data is computed upon inside an enclave, and plaintext never leaves it.

Proposed implementations~\cite{Biernacki2022Sequestered} suffer from duplication of computing resources in a specialized enclave, where the extra computing resources can only be used for PEC instructions, and PEC instructions do not have access to the full suite of CPU compute resources like multiple ALUs and vector units. SE's unit carries its own integer and floating-point datapath between its cipher engines, so every PEC operation needs duplicated compute resources.

In this paper, we introduce HPS-PEC, a novel architecture implementing sequestered encryption without requiring any extra computation resources apart from the ability to decrypt ciphertext and encrypt plaintext, and requiring minimal modifications to the core. We utilize existing core infrastructure, such as the renamer and $\mu Op$ expander to allow PEC instructions to share the same infrastructure as normal instructions while guaranteeing protections for sensitive data by utilizing renamer properties and hidden ``microarchitectural'' registers, and using the out-of-order mechanisms of the core to hide encryption and decryption latency and execute PEC instructions in parallel. Plaintext thus sits in plain sight in the register file, yet no architectural register can map to it.
With a 7-cycle QARMA-64, encrypted execution costs 3.41$\times$ native code and 3.31$\times$ data-oblivious plaintext code while only adding encryption and decryption units and minimally changing the core.

Previous hardware approaches to protecting data differ in where plaintext may exist inside the processor. Execute-only memory (XOM) tags registers with their owner so that the OS cannot read them, but trusts the whole core with plaintext~\cite{lie2000xom}. Data-oblivious ISA extensions confine confidential data to operands that do not leak it, without encrypting it~\cite{yu2019oisa}. Encrypted-execution designs, including SE, compute in a dedicated decrypt--compute--encrypt unit~\cite{chielle2022e3x, Biernacki2022Sequestered}, which can be dispatched like a native functional unit~\cite{Biernacki2022Sequestered, tan2023lowtrust}. HPS-PEC confines plaintext to hidden registers in an ordinary core instead, so it avoids the need for a trusted datapath.

\section{Design}

\begin{figure}[!t]
\centering
\includegraphics[width=0.78\linewidth]{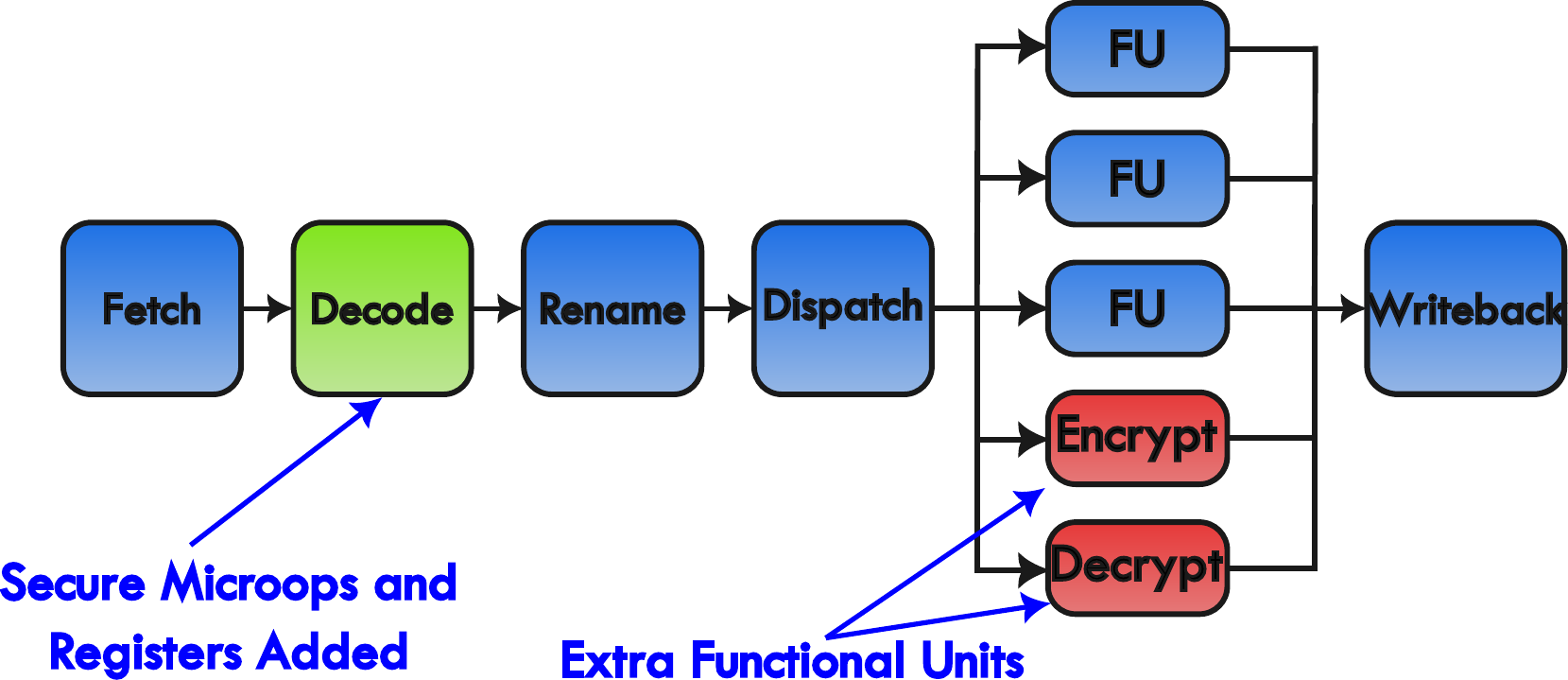}
\caption{HPS-PEC pipeline modifications. Decode expands each PEC instruction into micro-operations that use hidden microarchitectural registers, and encryption and decryption units are added alongside the existing functional units.}
\label{fig:our_solution}
\end{figure}

Our proposal is based on adding encrypt and decrypt function units alongside the other typical function units of a CPU and a set of PEC instructions. Register renaming is tasked to keep plaintext data completely private in physical registers, accessible only by PEC instructions. Like an enclave, this keeps plaintext out of reach of every instruction, in exchange for allowing PEC instructions to benefit from the latency-hiding abilities of an out-of-order micro-architecture. More specifically:
\begin{itemize}
    \item Encrypt and decrypt FUs are single-input/single-output, fixed-latency, pipelined units (one issue per cycle); Section~\ref{sec:crypto} estimates their latency. At least one of each type is needed; more can increase the encrypt/decrypt bandwidth at the expense of an increased area. 
    \item Similarly to SE~\cite{Biernacki2022Sequestered}, we utilize special instructions to denote PEC operations. Our implementation, however, is different. We utilize $\mu Op$ expansion and microarchitectural registers hidden from architectural state to implement these instructions.
\end{itemize}

\subsection{\texorpdfstring{\uop}{uOp} Expansion}

\begin{figure}[!t]
\centering
\includegraphics[width=0.75\linewidth]{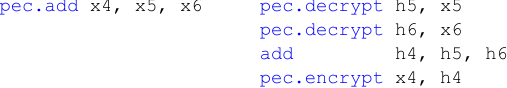}
\caption{Example of the expansion of PEC instructions into $\mu Ops$ showing how a regular add may be used to perform operations on hidden registers.}
\label{fig:uop_exp}
\end{figure}

When the CPU encounters a PEC instruction, it breaks it down into three to four micro operations as shown in \autoref{fig:uop_exp}: one to two decrypt the operands from normal registers into plaintext registers, one operates on those plaintext registers, and one encrypts the result back into a normal register.

This sequence acts like an enclave, but its operation runs on the core's own execution units, so no compute resources are duplicated. It also enables optimizations such as eliminating a decryption whose plaintext is still in a hidden register; we leave the evaluation of this optimization for future work.

\subsection{Microarchitectural Registers}
Microarchitectural registers are commonly used in CPU designs where instructions can break down into $\mu Ops$ which must not change architectural state, but may need to pass state between each other. One example is the gem5 implementation of RISC-V, where \texttt{Ureg0} is used when expanding certain $\mu Ops$. While these registers are treated by the microarchitecture as normal registers, and they are stored in the Physical Register File (PRF), they can only ever be accessed by the corresponding micro operations and may never be accessed by a normal instruction. This state is also normally never visible to $\mu Ops$ of a different macro instruction as they are typically written to before being read from.

This guarantee is upheld by the renamer, as the renamer never allows an instruction to access a physical register mapped to an incorrect architectural register, and while the physical register may be reused, it can only ever be mapped to a new register (architectural or microarchitectural) if it is written to, completely overwriting the previous state, so freed registers need no zeroing. As long as the renamer functionality holds, and the instructions that can access microarchitectural registers are well defined, normal instructions are never able to read the value in a hidden microarchitectural register. This allows us to store the results from the $\mu Op$ expansion in the physical register file with minor modifications to the microarchitecture while maintaining confidentiality.
Hidden registers have no instruction encoding: register fields name only \texttt{x0}--\texttt{x31}, and only the decoder's expansion of a PEC instruction emits hidden registers, so no software can read one. Each of the 32 hidden registers pairs with an architectural one (e.g. \texttt{h4} with \texttt{x4} in \autoref{fig:uop_exp}) and they are renamed similarly: every write takes a fresh physical register, so there are no false dependences, and the existing free list and rename checkpoints handle release and squash recovery. %

Because interrupts and squashes occur at macro instruction boundaries, transient plaintext values within microarchitectural registers are committed or squashed prior to a trap, requiring zero save/restore overhead for hidden register state. Microarchitectural registers are never used across different macro instructions as every macro instruction writes to every microarchitectural register (thus overwriting it) before consuming it. %

\subsection{Key Context}
Each process has a key held in dedicated registers inside the cipher units. On a context switch, the OS saves and restores only the key encrypted under an ephemeral key known only to the core, with a Message Authentication Code (MAC), so it can neither read nor alter it. As in SE~\cite{Biernacki2022Sequestered}, the data owner provisions the process key to the core through an authenticated key-exchange protocol built on public-key primitives such as Diffie--Hellman~\cite{diffiehellman} or RSA~\cite{RSA}.

\subsection{Threat Model}\label{sec:threat}
As in SE~\cite{Biernacki2022Sequestered}, the adversary controls all software, including the OS, but has no physical access, and programs are data-oblivious: their control flow and memory addresses do not depend on encrypted values. HPS-PEC guarantees that plaintext never reaches architectural state: registers and memory receive only ciphertext, produced by an encryption micro-operation. 
Unlike an enclave, it relies on the correctness of the renamer and the register file. Every program already depends on them, since a renamer that maps the wrong physical register corrupts ordinary execution with or without PEC, so HPS-PEC adds no correctness requirement of its own. Transient execution is in scope: even on a wrong path, only PEC micro-operations read hidden registers, and they produce only ciphertext. Port and unit contention reveals when PEC micro-operations run, which in data-oblivious code is independent of secrets, given that PEC micro-operations have operand-independent latency (no early-out division or subnormal slow paths) and raise no exceptions or floating-point flags. The 64-bit format reveals equality and permits replay (Section~\ref{sec:crypto}); we use it to measure the cost of the mechanism, not as a complete secure design.

\section{Methodology}

\subsection{System Configuration}

We extend GCC~\cite{gcc} with the PEC instructions, which mirror ALU/FPU operations, and gem5~\cite{gem5c1, gem5c2} to expand them into $\mu Ops$ and add the cipher units. The system configuration is shown in \autoref{tab:cpu_specs}; its window, scheduler, issue width and caches match Qualcomm's Oryon core. We test designs with and without loop unrolling and use the faster build.

\begin{table}[t]
\centering
\caption{Simulated CPU Microarchitecture Configuration}
\label{tab:cpu_specs}
\begin{tabular}{@{}ll@{}}
\toprule
\textbf{Component} & \textbf{Configuration Details} \\ \midrule
\textbf{Processor/ROB} & gem5 O3CPU, SE Mode, 2.5 GHz / 650-Entry \\
\textbf{Pipeline Width} & 8-wide Fetch/Decode/Commit, 14-wide Issue \\
\textbf{Instruction Queue} & 376-entry \\
\textbf{Register File} & 400 Int/FP registers \\
\textbf{Load/Store Queue} & 192-entry LQ, 144-entry SQ \\
\textbf{Branch Predictor/BTB} & Tournament / 4096 Entries \\
\textbf{Fetch} & 64B fetch buffer + 4$\times$64B instruction buffer \\ \midrule
\textbf{L1 I/D-Cache} & 192/96 KB, 6/6-way, 2/4-cycle, 8 MSHRs \\
\textbf{L2 Cache} & 12 MB, 12-way, 18-cycle, 64B line, 32 MSHRs \\ \midrule
\textbf{Main Memory} & DDR5-6400, 1 channel, 512 MB capacity \\
\bottomrule
\end{tabular}
\end{table}

\subsection{Cryptographic Algorithm}\label{sec:crypto}

For this work, we evaluate our design using a 64-bit cryptographic algorithm, namely QARMA-64~\cite{QARMA}. While our approach can generalize to 128 bits or more in different ways, keeping the cryptographic algorithm at 64 bits allows us to quantify the overheads of the approach itself, rather than the overheads of doubling the register width and thus all memory accesses. This choice has a security cost: a 64-bit block leaves no room for per-value randomness, so equal plaintexts produce equal ciphertexts, revealing, for example, which input an encrypted select returns, and observed ciphertexts can be replayed undetected. Closing these gaps needs per-value randomness, as in SE's 128-bit format~\cite{Biernacki2022Sequestered}; that doubles register and memory width and needs optimizations of its own, which we leave to future work. We use QARMA$_7$-64-$\sigma_0$ ($r=7$, the recommended parameter: seven forward and seven backward rounds around a central construction). Our implementation reproduces the published test vectors in RTL simulation and, placed and routed with OpenROAD~\cite{ajayi2019openroad} in the ASAP7 7\,nm library~\cite{clark2016asap7}, meets 2.5\,GHz with seven pipeline stages, so encryption and decryption take 7 pipelined cycles (Section~\ref{sec:sensitivity} varies this from 1 to 10).

\subsection{Changes to VIP-Bench}
VIP-Bench~\cite{biernacki2021vip} builds each benchmark as native code (NA), as data-oblivious code on plaintext (DO), and as encrypted code (ENC). To avoid branching on sensitive data, DO mode requires conditional moves. RISC-V's Zicond extension needs three instructions for a conditional move, whereas our encrypted select is one instruction. To ensure a fair comparison, we add a one instruction conditional move where the condition register is overwritten by the result. 

As shipped, VIP-Bench's sources do not always measure what the suite intends, so we made six changes and left everything else as upstream.  (1)~Timed regions included printing, the decryptions it needs, input encryption or allocation, and some were too short to time; we moved such work out and repeat short kernels 8 to 1024 times. (2)~Short-circuit \texttt{\&\&} and \texttt{||} branched on secret values, so the DO build was not oblivious: it leaked, and skipped work that ENC performs; we made them branch-free.  (3)~A logic error in \texttt{gcd-list} skipped half of its divisions.  (4)~The host cannot encrypt values, as it does not hold the key; the benchmarks' encryption of their own inputs only stands in for their delivery.  Constants and public values, which were encrypted inside the timed region, therefore enter encrypted--plaintext operations as plaintext, and a constant that must be ciphertext is encrypted once with the inputs.  (5)~Compiler artifacts that gave the NA or DO build of four kernels extra taken branches or memory reloads absent from ENC were fixed.
(6)~Many kernels were needlessly serial or redundant: reductions used a single accumulator, sorts and pair loops revisited elements already in place, and some expressions were computed twice, which GCC merges for plaintext (NA, DO) but cannot for encrypted operations; we split reductions four ways and removed the redundant work.  Every change applies to all modes that run the affected code, and every benchmark still produces its reference output. Overall, these fixes raise the measured cost of encryption: with a 1-cycle cipher, the unmodified sources give ENC 1.45$\times$ NA instead of 2.06$\times$.

\section{Results}

\subsection{Performance}

\begin{figure*}[!t]
\centering
\includegraphics[width=\linewidth]{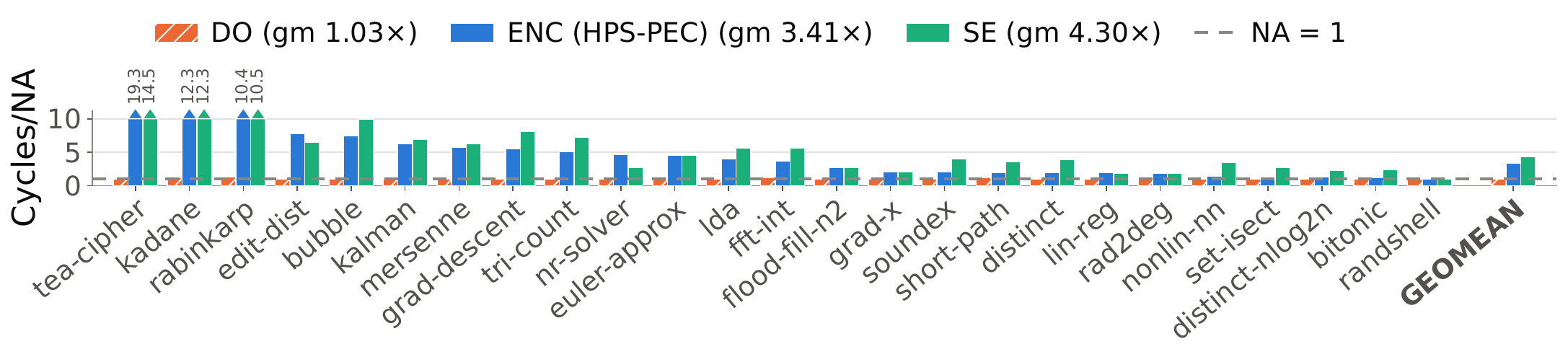}
\caption{Execution time of the data-oblivious (DO) and encrypted (ENC) builds
and the fixed-function SE unit, normalized to native code (NA${}=1$), with a
7-cycle cipher; HPS-PEC has one encryption and two decryption units. Benchmarks are sorted by ENC overhead and the
legend gives geometric means. Bars above 10$\times$ end in a triangle labeled with their value. }
\label{fig:performance}
\end{figure*}

\label{sec:performance}
\autoref{fig:performance} shows the overhead of PEC: the algorithmic cost of data-oblivious code (DO/NA), a lower bound that no PEC hardware can beat, and the cost of encryption (ENC/DO), which HPS-PEC adds: 1.03$\times$ and 3.31$\times$ respectively, for 3.41$\times$ in total (geomean~\cite{Nematallah2026One}). Encryption costs least in \texttt{randshell-sort} (1.0$\times$) and \texttt{bitonic-sort} (1.2$\times$), where the out-of-order core hides it, and most in \texttt{tea-cipher} (19.3$\times$), \texttt{kadane} and \texttt{rabinkarp-search}, whose chains of dependent encrypted operations expose every encryption and decryption.

\autoref{fig:performance} also shows SE~\cite{Biernacki2022Sequestered}: a pipelined functional unit whose commands are issued like any other FU. A command decrypts its operands, computes and re-encrypts the output, taking the encryption and decryption latencies plus the operation's own latency. As in SE's evaluation, the unit has one instance of each component. With code generation matched to ours, SE is 1.26$\times$ slower than HPS-PEC (geomean), slower on 16 benchmarks and faster on 5. HPS-PEC gains most on select-heavy code (up to 2.48$\times$, \texttt{nonlinear-nn}), where a select is one ordinary micro-operation on hidden registers in HPS-PEC but a full decrypt--compute--encrypt command in SE; SE wins on long dependent chains (\texttt{nr-solver}, 0.59$\times$), where four registers per PEC instruction exhaust HPS-PEC's register file.

\subsection{Sensitivity to the Cipher and the Core}\label{sec:sensitivity}

\autoref{tab:sensitivity} varies one parameter at a time. Seven benchmarks, such as \texttt{grad-x} and \texttt{soundex}, do not slow down from 1 to 10 cycles because the out-of-order core overlaps the cipher with independent work, while dependent chains (\texttt{tea-cipher}, \texttt{kadane}) slow down up to 6.7$\times$. Latency hurts SE too, which pays $2L_c$ per command. Most PEC instructions decrypt two operands, so a single decryption unit costs HPS-PEC 16\% (up to 1.5$\times$ in \texttt{soundex}). Each PEC instruction occupies 3--4 window entries instead of SE's one, so smaller cores cost HPS-PEC more.

\begin{table}[t]
\centering
\caption{Sensitivity (geomean over 25 benchmarks); ENC and SE over native code, and SE over HPS-PEC (${>}1$: HPS-PEC faster).}
\label{tab:sensitivity}
\begin{tabular}{@{}l@{\hspace{4pt}}c@{\hspace{4pt}}c@{\hspace{4pt}}c@{}}
\toprule
\textbf{Varied} & \textbf{ENC/NA} & \textbf{SE/NA} & \textbf{SE/ENC} \\ \midrule
$L_c$ 1/7/10 cycles & 2.06/3.41/4.10 & 3.34/4.30/4.77 & 1.62/1.26/1.16 \\
ROB 256/512/650 & 4.43/3.52/3.41 & 4.24/4.27/4.30 & 0.96/1.21/1.26 \\
Scheduler 160/376 & 3.33/3.41 & 4.25/4.30 & 1.27/1.26 \\
Units 1E1D/1E2D/2E4D & 3.94/3.41/3.30 & -- & -- \\
\bottomrule
\end{tabular}
\end{table}

\subsection{Area and Energy}\label{sec:area}
We placed and routed all units with OpenROAD in ASAP7 at a 2.5\,GHz target: a cipher unit (encryptor or decryptor, the same circuit) takes 1699\,$\mu$m$^2$, an RV64 integer unit (ALU, multiplier, divider) 6039\,$\mu$m$^2$, and an FP32/FP64 FPU (CVFPU~\cite{mach2021fpnew}, with its FMA multiplier pipelined) 7762\,$\mu$m$^2$; the integer unit and FPU do not close 2.5\,GHz in the open flow, so their areas are lower bounds. SE as modeled (one encryptor, one decryptor) thus takes at least 17\,199\,$\mu$m$^2$, and the three cipher units of HPS-PEC 5097\,$\mu$m$^2$, under 30\% of the area of SE. 

Whole-run energy was computed with the core from McPAT-7nm~\cite{Zhai2023McPAT}, DRAM from gem5, and the units from gate-level simulation of the routed netlists (calibrated to McPAT's integer ALU). SE uses 0.99$\times$ the energy of HPS-PEC (geomean), from 0.64$\times$ in \texttt{nr-solver} to 1.41$\times$ in \texttt{bubble-sort}, as HPS-PEC's extra micro-operations cost about as much core energy as SE's longer runtime. HPS-PEC itself uses 2.30$\times$ the energy of native code (geomean; 1.05$\times$ for DO), of which its cipher units account for under 0.3\%.

\section{Conclusion}

We presented HPS-PEC, a PEC design that keeps plaintext in hidden microarchitectural registers, protected by register renaming, and breaks PEC instructions into ordinary $\mu Ops$, sharing the core's functional units and out-of-order scheduling instead of an enclave. On VIP-Bench it costs 3.41$\times$ native and 3.31$\times$ data-oblivious code with 64-bit QARMA at 7 cycles, 1.26$\times$ faster than an SE unit while taking up under a third of its dedicated area; its 3--4 micro-operations per PEC instruction are its main cost. Our design enables eliminating redundant decryptions and supporting 128-bit ciphers, both left for future work.

\bibliographystyle{IEEEtran}
\bibliography{refs,bstctl}

\begin{thebibliography}{10}
\providecommand{\url}[1]{#1}
\csname url@samestyle\endcsname
\providecommand{\newblock}{\relax}
\providecommand{\bibinfo}[2]{#2}
\providecommand{\BIBentrySTDinterwordspacing}{\spaceskip=0pt\relax}
\providecommand{\BIBentryALTinterwordstretchfactor}{4}
\providecommand{\BIBentryALTinterwordspacing}{\spaceskip=\fontdimen2\font plus
\BIBentryALTinterwordstretchfactor\fontdimen3\font minus \fontdimen4\font\relax}
\providecommand{\BIBforeignlanguage}[2]{{%
\expandafter\ifx\csname l@#1\endcsname\relax
\typeout{** WARNING: IEEEtran.bst: No hyphenation pattern has been}%
\typeout{** loaded for the language `#1'. Using the pattern for}%
\typeout{** the default language instead.}%
\else
\language=\csname l@#1\endcsname
\fi
#2}}
\providecommand{\BIBdecl}{\relax}
\BIBdecl

\bibitem{biernacki2021vip}
L.~Biernacki \emph{et~al.}, ``{VIP-Bench}: A benchmark suite for evaluating privacy-enhanced computation frameworks,'' in \emph{SEED}, 2021, pp. 139--149.

\bibitem{Biernacki2022Sequestered}
L.~Biernacki \emph{et~al.}, ``Sequestered encryption: A hardware technique for comprehensive data privacy,'' in \emph{2022 IEEE International Symposium on Secure and Private Execution Environment Design (SEED)}, 2022, pp. 73--84.

\bibitem{lie2000xom}
D.~Lie \emph{et~al.}, ``Architectural support for copy and tamper resistant software,'' in \emph{ASPLOS}.\hskip 1em plus 0.5em minus 0.4em\relax ACM, 2000, pp. 168--177.

\bibitem{yu2019oisa}
J.~Yu \emph{et~al.}, ``Data oblivious {ISA} extensions for side channel-resistant and high performance computing,'' in \emph{NDSS}.\hskip 1em plus 0.5em minus 0.4em\relax Internet Society, 2019.

\bibitem{chielle2022e3x}
E.~Chielle \emph{et~al.}, ``{E3X}: Encrypt-everything-everywhere {ISA} extensions for private computation,'' \emph{IEEE Trans. Dependable Secure Comput.}, vol.~19, no.~2, pp. 848--861, 2022.

\bibitem{tan2023lowtrust}
Q.~Tan \emph{et~al.}, ``Security verification of low-trust architectures,'' in \emph{ACM CCS}.\hskip 1em plus 0.5em minus 0.4em\relax ACM, 2023, pp. 945--959.

\bibitem{diffiehellman}
W.~Diffie and M.~Hellman, ``New directions in cryptography,'' \emph{IEEE Transactions on Information Theory}, vol.~22, no.~6, pp. 644--654, 1976.

\bibitem{RSA}
\BIBentryALTinterwordspacing
R.~L. Rivest, A.~Shamir, and L.~Adleman, ``A method for obtaining digital signatures and public-key cryptosystems,'' \emph{Commun. ACM}, vol.~21, no.~2, pp. 120--126, Feb. 1978. [Online]. Available: \url{https://doi.org/10.1145/359340.359342}
\BIBentrySTDinterwordspacing

\bibitem{gcc}
{Free Software Foundation}, \emph{GCC online documentation}, Boston, MA, USA, August 2025, {Accessed}: Apr. 7, 2026.

\bibitem{gem5c1}
\BIBentryALTinterwordspacing
J.~Lowe-Power \emph{et~al.}, ``The gem5 simulator: Version 20.0+,'' 2020. [Online]. Available: \url{https://arxiv.org/abs/2007.03152}
\BIBentrySTDinterwordspacing

\bibitem{gem5c2}
\BIBentryALTinterwordspacing
N.~Binkert \emph{et~al.}, ``The gem5 simulator,'' \emph{SIGARCH Comput. Archit. News}, vol.~39, no.~2, pp. 1--7, Aug. 2011. [Online]. Available: \url{https://doi.org/10.1145/2024716.2024718}
\BIBentrySTDinterwordspacing

\bibitem{QARMA}
\BIBentryALTinterwordspacing
R.~Avanzi, ``The {QARMA} block cipher family -- almost {MDS} matrices over rings with zero divisors, nearly symmetric {Even-Mansour} constructions with non-involutory central rounds, and search heuristics for low-latency {S-Boxes},'' Cryptology {ePrint} Archive, Paper 2016/444, 2016. [Online]. Available: \url{https://eprint.iacr.org/2016/444}
\BIBentrySTDinterwordspacing

\bibitem{ajayi2019openroad}
T.~Ajayi \emph{et~al.}, ``Toward an open-source digital flow: First learnings from the {OpenROAD} project,'' in \emph{Design Automation Conference (DAC)}, 2019.

\bibitem{clark2016asap7}
L.~T. Clark \emph{et~al.}, ``{ASAP7}: A 7-nm {finFET} predictive process design kit,'' \emph{Microelectronics Journal}, vol.~53, pp. 105--115, 2016.

\bibitem{Nematallah2026One}
A.~Nematallah, ``One mean can’t rule them all,'' \emph{Computer}, vol.~59, no.~2, pp. 49--58, 2026.

\bibitem{mach2021fpnew}
S.~Mach \emph{et~al.}, ``{FPnew}: An open-source multiformat floating-point unit architecture for energy-proportional transprecision computing,'' \emph{IEEE Transactions on Very Large Scale Integration (VLSI) Systems}, vol.~29, no.~4, pp. 774--787, 2021.

\bibitem{Zhai2023McPAT}
J.~Zhai \emph{et~al.}, ``{McPAT-Calib}: A {RISC-V} {BOOM} microarchitecture power modeling framework,'' \emph{IEEE Transactions on Computer-Aided Design of Integrated Circuits and Systems}, vol.~42, no.~1, pp. 243--256, 2023.

\end{thebibliography}

\vfill

\end{document}